\documentclass[10pt]{iopart}
\usepackage{iopams,mathptmx}
\usepackage{graphicx,enumerate}
\usepackage[colorlinks=true,urlcolor=blue,citecolor=blue,linkcolor=blue]{hyperref}
\usepackage[normalem]{ulem}

\begin{document}

\title[Quasi\-probability distributions]{Quasi\-probability distributions for quantum-optical coherence and beyond}

\author{J. Sperling}
\address{Integrated Quantum Optics Group, Applied Physics, University of Paderborn, Warburger Stra\ss{}e 100, 33098 Paderborn, Germany}
\ead{jan.sperling@uni-paderborn.de}

\author{W. Vogel}
\address{Institut f\"ur Physik, Universit\"at Rostock, Albert-Einstein-Stra\ss{}e 23, 18059 Rostock, Germany}
\ead{werner.vogel@uni-rostock.de}

\begin{abstract}
	We study the quasi\-probability representation of quantum light, as introduced by Glauber and Sudarshan, for the unified characterization of quantum phenomena.
	We begin with reviewing the past and current impact of this technique.
	Regularization and convolution methods are specifically considered since they are accessible in experiments.
	We further discuss more general quantum systems for which the concept of negative probabilities can be generalized, being highly relevant for quantum information science.
	For analyzing quantum superpositions, we apply recently developed approaches to visualize quantum coherence of states via negative quasi\-probability representations, including regularized quasi\-probabilities for light and more general quantum correlated systems.
\end{abstract}

\begin{indented}
\item[] \textit{Dated}: \today
\item[]
\item[] \textit{Keywords}: Quasi\-probability distributions, quantum optics, quantum coherence
\end{indented}

\ioptwocol

\section{Introduction}

	At the beginning of the 20th century, Planck \cite{P01} and Einstein \cite{E05} offered new explanations for experimental observations that, in retrospect, can be acknowledged as the beginning of the field of quantum optics, an area of science devoted to the study of the quantum-mechanical properties of electromagnetic radiation \cite{MW95,VW06,A12}.
	Later, Wigner constructed a phase-space probability representation for wave functions \cite{W32,M49} which, however, can become negative in certain scenarios.
	Because of this and other incompatibilities of quantum theory with classically preconceived concepts of nature, quantum physics was widely considered as a counterintuitive theory, despite its remarkable success in predicting the outcomes of experiments with exceptionally high accuracy.

	In their seminal works \cite{G63,S63}, Glauber and Sudarshan formulated an, at the time, new methodology by establishing a framework of phase-space quasi\-probability distributions in quantum optics.
	The failure to interpret this specific distribution in terms of classical statistical models of light defines the very concept of a nonclassical state \cite{TG65,L86}. It also constitutes the first self-consistent and universally applicable technique to understand quantum phenomena by extending the classical concept of a phase space to the quantum realm.
	It is noteworthy that the concepts of quantum-optical coherence developed by Glauber and his coworkers extend to nonlinear quantum-optical processes as well \cite{FG71,MG67b,MG67a}.
	For his overall contributions to the theory of coherence of quantum light, as well as about one hundred years after the notion of light quanta was proposed \cite{G06,G07b}, Roy J. Glauber was awarded with the Nobel prize in 2005; see \cite{G07a} for his own selection of related papers.

	In this contribution, we study quasi\-probability representations for a unified understanding of quantum phenomena.
	For this purpose, this paper is subdivided into two parts.
	First, in Section \ref{sec:Review}, we present a more comprehensive review that demonstrates how quasi\-probability distributions became a standard tool for analyzing quantum effects in theory and experiment.
	We discuss how this concept of quasi\-probabilities inspired a variety of research directions that further improved existing approaches and even extends to other concepts of quantumness for more general systems and applications in quantum information theory.
	Second, in Section \ref{sec:Coherence}, we apply some recently developed techniques to represent quantum interference using quasi\-probabilities.
	In particular, the impact of quantum superposition on telling patterns in regularized and experimentally accessible quasi\-probabilities is investigated.
	We conclude in Section \ref{sec:Summay}.

	In summary, we review the impact and extend of the methodology of quasi\-probabilities, widely used to discern quantum interference from classical ones in optical systems and beyond.
	By doing so, we are able to provide a generally applicable and convolution-based framework which enables us to describe arbitrary nonclassical superposition states.
	Furthermore, this technique renders it possible to theoretically and experimentally study the origin of coherence phenomena, potentially having a classical or quantum interpretation, via intuitively accessible phase-space representations.

\section{Quasi\-probability representations in quantum physics}\label{sec:Review}

\subsection{Phase-space methods in quantum optics}

	For a single-mode light field, the Glauber-Sudarshan (GS) representation of a density operator reads \cite{G63,S63}
	\begin{equation}
		\label{eq:GSrepresentation}
		\hat\rho=\int \rmd^2\alpha\,P(\alpha)|\alpha\rangle\langle\alpha|,
	\end{equation}
	where $P$ is often referred to as GS function or distribution, $|\alpha\rangle$ defines a pure coherent state with the complex amplitude $\alpha$, and $\rmd^2\alpha=\rmd\mathrm{Re}\,\alpha\,\rmd\mathrm{Im}\,\alpha$.
	Note that all functions with a complex argument are understood here and in the following as functions of two real arguments, $\mathrm{Re}\,\alpha$ and $\mathrm{Im}\,\alpha$.
	Relatively early in the history of quantum physics, the coherent state has been found to resemble the behavior of a macroscopic (\textit{i.e.} classical) harmonic oscillator most closely \cite{S26}.
	In fact, much later, it was shown that the only pure states that are classical in a harmonic oscillator system are those coherent states \cite{H85}, which in itself was based on definition \eref{eq:GSrepresentation}.

	More generally, as long as a GS distribution can be found for a given state that can be interpreted as a nonnegative probability density, $P\geq0$, the corresponding quantum state is defined to be classical.
	Conversely, if the state cannot be represented in this form, the notion of a nonclassical state applies \cite{TG65,L86}.
	Interestingly, it is still possible to expand such nonclassical states in the form of \eref{eq:GSrepresentation} \cite{S63,TG65} which then, however, means that $P$ does no longer satisfy the nonnegativity premise, \textit{i.e.} $P\ngeq0$.
	To date, this concept of a nonclassical state forms the basis for discerning quantum optics from classical statistical optics.
	See also \cite{K13} for a comparison of classical and quantum-mechanical phase-space representations.

	Although the formal definition of nonclassicality presents one of the most fundamental concepts of quantum optics, the GS $P$ function also exhibits some major drawbacks.
	In many scenarios, the $P$ function is a highly singular distribution which cannot be represented through a well-behaved function.
	Still, approximations have been conceived that enable us to approach this distribution with square-integrable \cite{KMC65} and even smooth \cite{K66} functions.
	Within the space of distributions, tempered distributions can be also used to approximate general $P$ functions \cite{MS65}.
	This is relevant as, in actuality, GS functions can have singularities which are even higher than expected from previously studied families of distributions.
	In fact, the highest singularity of GS functions which can be achieved is of an exponential order, $\rme^{-\partial_{\alpha}\partial_{\alpha^\ast}/2}\delta(\mathrm{Re}\,\alpha)\delta(\mathrm{Im}\,\alpha)$ \cite{S16}, where $\delta$ denotes the one-dimensional Dirac-$\delta$ distribution and $\partial_x$ the derivative with respect to an argument $x$.
	In addition, there are other limitations to physical states which concern the existence of higher-order moments \cite{C65} such that, for example, the total energy of a state is bounded.

	A less frequently considered problem is that the $P$ function of a state is ambiguous \cite{BV87}.
	This means that a single state $\hat\rho$ can be represented via different distributions.
	For example, both $P(\alpha)=\rme^{\bar n\partial_{\alpha}\partial_{\alpha^\ast}}\delta(\mathrm{Re}\,\alpha)\delta(\mathrm{Im}\,\alpha)$ and $P(\alpha)=\rme^{-|\alpha|^2\bar n}/(\pi\bar n)$ represent a thermal state for a mean photon number $\bar n>0$ \cite{S16,G63a}.
	Note that the former distribution appears to be highly singular while the latter is a smooth and nonnegative function;
	but both can be shown to act on test functions in the same manner \cite{S16}.
	What complicates matters even further is that every classical state has a nonclassical state in its neighbourhood \cite{W04}, which, in practice, makes it impossible to unambiguously certify that a state is classical within a finite error margin.
	Conversely, some more sophisticated approaches also demonstrate that each state can be represented through a regular and positive phase-space distribution which, however, is achieved by extending the argument $\alpha$ of $P$ to two complex variables for a single-mode state \cite{DG80,CAS94}, thus not representing the same phase space.

	In order to overcome many of the previously mentioned challenges, and for a more practical approach to the essential definition of nonclassicality, the GS $P$ function may be convoluted with an appropriate kernel $K$ \cite{CG69a,AW70b},
	\begin{equation}
		\label{eq:Convolution}
		P^{K}(\alpha)=\int \rmd^2\tilde\alpha\,K(\alpha,\tilde\alpha)P(\tilde\alpha),
	\end{equation}
	leading to the modified phase-space function $P^{K}$.
	For example, Husimi's phase-space function \cite{H40,K65}, given by
	$Q(\alpha)=\langle\alpha|\hat\rho|\alpha\rangle/\pi
	=\int \rmd^2\tilde\alpha\, [|\langle\alpha|\tilde\alpha\rangle|^2/\pi] P(\tilde\alpha)$,
	can be interpreted in terms of \eref{eq:Convolution} when using the Gaussian kernel $K(\alpha,\tilde\alpha)=|\langle\alpha|\tilde\alpha\rangle|^2/\pi=\rme^{-|\alpha-\tilde\alpha|^2}/\pi$.
	While this function is, by construction, always nonnegative, characteristic features of this distribution enable us to infer nonclassical properties of the state \cite{LB95,WV00,MS04}.
	Taking a Gaussian kernel with half the width as used for the $Q$ function, we get the Wigner function \cite{W32,M49}.
	For more general widths, one obtains the encompassing family of so-called $s$-parametrized phase-space distributions \cite{CG69a,AW70b}.
	As a Gaussian kernel $K$ is nonnegative, distributions $P^K$ can exhibit negativities only if the the original $P$ function describes a nonclassical state.
	However, the inverse is not true as nonclassical states can have a nonnegative Wigner function, \textit{e.g.} for squeezed states.

\begin{figure}
	\includegraphics[width=\columnwidth]{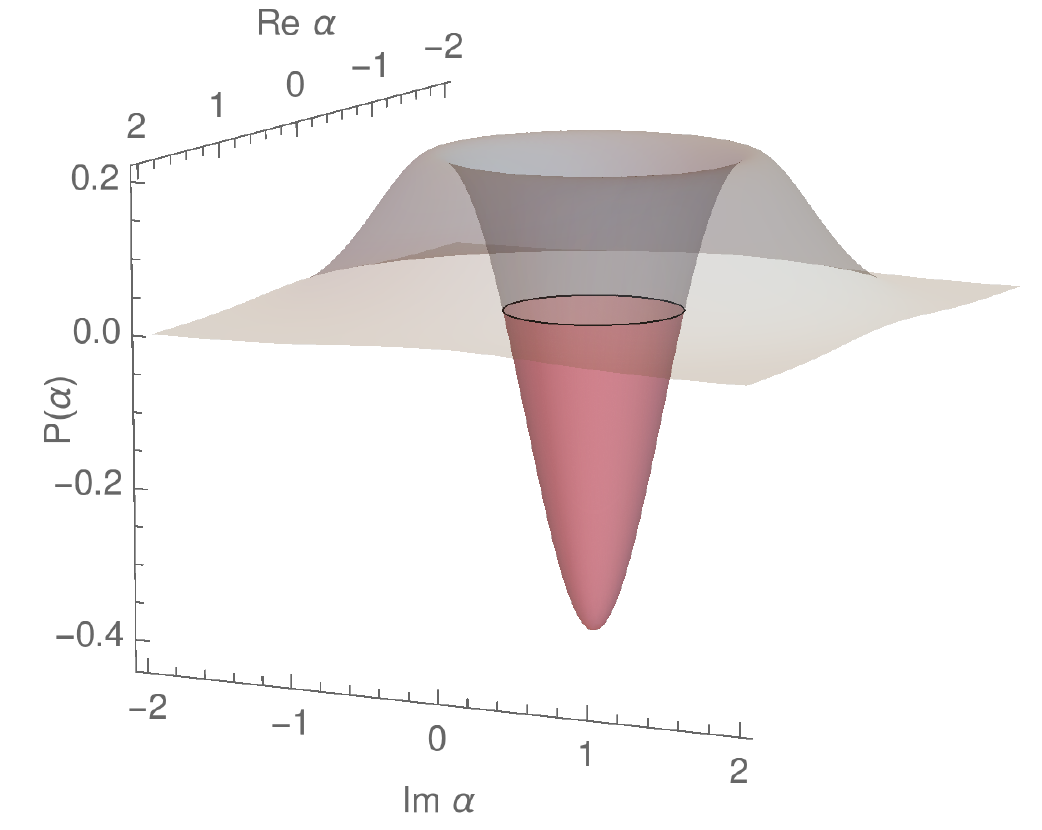}
	\caption{
		The GS $P$ function for a nonclassical single-photon-added thermal state \cite{AT92} is shown for parameters that correspond to the experimental conditions reported in \cite{KVPZB08}.
	}\label{fig:SPATS}
\end{figure}

	Nevertheless, Wigner's phase-space representation turned out to be a versatile instrument for many experimental applications as it does not yield singular distributions.
	Some of the remarkable examples which have been experimentally probed via the Wigner function approach are squeezed states \cite{SBRF93, HSRHMSS16}, Fock states (\textit{e.g.} single photons) \cite{LMKMIW96,LHABMS01,ZPB07,DDSBBRH08,LCGS10}, Schr\"odinger's cat states \cite{MMKW96,OTLG06,VKLNFGMDS13,MHLJFL14,SUPRFL17}, and even more complex quantum states \cite{SWM95,DEWKDKCM13,BFS17}.
	In fact, the first experimental reconstruction of a true GS distribution was achieved only relatively recently \cite{KVPZB08} and was based on states theoretically proposed in \cite{AT92} with a regular $P$ function;
	see also Figure \ref{fig:SPATS}.

	As singularities pose a challenging problem when dealing with experimental data, and $s$-parametrized functions do not necessarily include negative contributions for nonclassical states, alternative approaches were needed.
	A proper regularization of the $P$ function has been constructed relatively early \cite{K66}, using specific convolutions of the form \eref{eq:Convolution}.
	This construction, however, can lead to negativities in the resulting distribution even for classical states.
	To achieve both goals---\textit{i.e.} nonnegativity for classical states and regularity---at the same time, more elaborate techniques were required.
	Specifically, non-Gaussian kernels $K$ can be used \cite{AW70b}. 
	When constructed carefully with proper constraints \cite{KV10}, the regularized quasi\-probability distributions $P^K$ exhibits negativities for any nonclassical light but never for classical light.
	Hence, the resulting distributions refer to as nonclassicality quasi\-probabilities.
	Their construction requires to find a sufficiently smooth kernel $K$ which, most importantly, is nonnegative for all arguments \cite{KV10,KV14}.
	Applying this approach to experimental data even reveals the nonclassicality of squeezed states via regular quasi\-probabilities with negativities \cite{KVHS11,ASVKMH15}, which is impossible using $s$-parametrized distributions.
	It is also noteworthy that the regularization of the $P$ function can be implemented as a technique of state engineering, which prepares out of any nonclassical state of light a state with a regular $P$ function \cite{KV18a}.

	It is also worth mentioning that a deconvolution---\textit{i.e.} the inverse operation to \eref{eq:Convolution}---might be useful to extract the information about the GS distribution from the convoluted one \cite{AW70b,KV10}.
	When the kernel is invertible, a convolution with the inverse kernel $K^{-1}$, satisfying
	$\int \rmd^2\tilde\alpha\, K^{-1}(\alpha,\tilde\alpha)K(\tilde\alpha,\alpha')=\delta(\mathrm{Re}\,\alpha-\mathrm{Re}\,\alpha')\delta(\mathrm{Im}\,\alpha-\mathrm{Im}\,\alpha')$, renders it possible to retrieve $P$ from $P^K$ via the relation $(P^K)^{K^{-1}}=P$.

	In quantum physics, the state itself is meaningless when measurements are not taken into account.
	Using the GS formalism, we can also write the quantum-mechanical expectation value of an observable $\hat M$ as
	\begin{equation}\eqalign{
		\tr(\hat\rho\hat M)
		&= \pi\int \rmd^2\alpha\, P_{\hat\rho}(\alpha)Q_{\hat M}(\alpha)
		\\
		&= \pi\int \rmd^2\tilde\alpha\, P^K_{\hat\rho}(\tilde\alpha)Q^{K^{-1}}_{\hat M}(\tilde\alpha),
	}\end{equation}
	where $P^{K}_{\hat X}$ and $Q^K_{\hat X}$ are the GS function and the Husimi function, respectively, for an operator $\hat X$ and an integral transformation kernel $K$.
	This then also implies the existence of a GS-type representation for an observable similarly to the state's representation \eref{eq:GSrepresentation} when using an invertible kernel \cite{C65,M67}.
	Consequently, the characteristic quantum features of detectors, rather than states, can be analyzed experimentally using phase-space representation, such as demonstrated for a single-photon counter \cite{LFCPSREPW09} and a homodyne detector \cite{GZBP17}.

	Moreover, a specific phase-space representation can be connected with certain orderings of noncommuting operators \cite{AW68b}.
	For example, the $s$-parametrized quasi\-probability distributions render a continuous transition from normal ordering (\textit{i.e.} annihilation operators after creation operators for $P$ functions) over symmetric ordering (for Wigner functions) to antinormal ordering (\textit{i.e.} annihilation operators before creation operators for $Q$ functions) \cite{CG69b,AW70a}.
	For instance, correlation functions, an essential tool for characterizing light, are expressed in normally ordered photon-number correlations \cite{G63b}, relating to the definition of nonclassicality via the GS distribution.
	Similarly, one can order the quadrature operator together with its conjugate momentum, defining the notion of standard and antistandard ordering \cite{C66,AW70a,AW68a}.
	Likewise, moments of regularized GS functions can be expressed via generalized nonclassicality orderings \cite{RKV12}.

	Furthermore, and beyond Gaussian convolution kernels, the density operator $\hat\rho$ can be also expanded as $\hat\rho=\int \rmd^2\alpha\,P^{K}(\alpha)\hat\Delta^{K^{-1}}(\alpha)$, using
	\begin{equation}
		\label{eq:Quasistates}
		\hat\Delta^{K^{-1}}(\alpha)=\int \rmd^2\tilde\alpha\,K^{-1}(\tilde\alpha,\alpha)|\tilde\alpha\rangle\langle\tilde\alpha|,
	\end{equation}
	which is similar to the GS decomposition \eref{eq:GSrepresentation}.
	For the specific case of Gaussian kernels $K$, $\hat\Delta^{K^{-1}}$ represents the previously mentioned operator ordering of an exponential of the displaced photon-number operator \cite{CG69b,AW70a}.
	In general, however, the operator in Equation \eref{eq:Quasistates} is not positive semi-definite, \textit{i.e.} $\hat\Delta^{K^{-1}}(\alpha)\ngeq0$, which elevates the concept of quasi\-probabilities to the analogous notion of a quasistate \cite{SW18b}.
	This recently conceived concept is useful as it enables us, for example, to resolve certain reconstruction problems in quantum optics \cite{KSVS18,SW18b}.

	Interestingly, the convolution with a kernel can be also associated with the action of a process of quantum channel \cite{M07,LKKFSL08,RKVGZB13}, \textit{e.g.} mapping a coherent state by the operation $\Lambda$ as $\Lambda(|\alpha\rangle\langle\alpha|)=\int \rmd^2\tilde\alpha\, K(\tilde\alpha,\alpha)|\tilde\alpha\rangle\langle\tilde\alpha|$.
	Equivalently, this means for mixed states that an input GS distribution, $P_\mathrm{in}$, is mapped onto a resulting one, $P_\mathrm{out}$, via the relation $P_\mathrm{out}(\tilde\alpha)=\int \rmd^2\alpha\,K(\tilde\alpha,\alpha)P_\mathrm{in}(\alpha)$.
	As for any classical channel $K\geq0$ holds true by definition, one gets an experimentally accessible identifier of the nonclassicality of processes in this manner as well \cite{RKVGZB13}.

	More generally, the evolution of a system can be analyzed in terms of the propagation of quasi\-probabilities in time \cite{G66,AW68a,C76}.
	Again, a convolution with suitable kernels enables us to consider generalized phase-space distributions \cite{AW70c}.
	It then turns out that the nonclassicality of a dynamical process is additionally connected to time ordering \cite{AW70c}---an effect which occurs for operators that commute for equal time but become non-commuting when correlated at different points in time.
	Consequently, the notion of the GS function can be generalized to space-time-based functionals by including time ordering \cite{V08}.
	A subsequent regularization of the resulting singular phase-space functionals can be performed as well \cite{KVS17}.

	The above treatments, including convolution techniques, can be equally and straightforwardly generalized to multiple optical modes \cite{TG66,ASV13}.
	In such a case, quantum correlations between optical modes are not only accessible via negativities in a joint quasi\-probability desciption, $P(\alpha_1,\alpha_2)$, but also accessible via conditional quasi\-probabilities, $P(\alpha_1|\alpha_2)$; see \cite{M07} for an early work connecting the usefulness of such an approach to quantum gates.
	Partial marginals of multimode phase-space distributions can also serve as a convenient way to describe the nonclassicality accessible in given interference-based measurement scenarios \cite{SVA16}.
	Note that even in the single-mode case, marginal distributions can be conclusive for identifying nonclassical states of light in theory and experiments \cite{A93a,KVCBAP12,PLLSZZZKN17}.

	Beyond systems of light, any contiuous-variable system that resembles a harmonic oscillator can be analyzed using GS quasi\-probabilities; for example, see \cite{DGW81,VR89,WR91} for applications.
	Quasi\-probabilities are also usful for assessing the importance of nonclassicality when performing modern quantum information protocols \cite{RRC16,SLR17}, bridging the gap between well-established concepts and recent innovations.
	This point is explored in more detail in the following.
	For this purpose, it is worth mentioning that for instance, the notion of non-Gaussianity can be evaluated when replacing the pure coherent states in definition \eref{eq:GSrepresentation} by general Gaussian states as the classical reference \cite{BV87,FM11,KV18,SW18b}.

\subsection{Quasi\-probabilities for other forms of quantumness}

	Because of the intuitive nature to visualize quantum effects via quasi\-probabilities, the desire for a generalization of classically impossible, negative probabilities to other fields of research---being only remotely related to quantum optics---is not surprising.
	For instance, quasi\-probabilities can be assigned to probe for contextuality \cite{S08,FE09}, a phenomenon in which the measurement context becomes relevant, such as measurements with noncommuting observables.
	Similarly, nonlocality \cite{LJJ09} (\textit{i.e.} Bell-type correlations) and quantum phenomena related to Leggett-Garg inequalities \cite{Ref2e,JBLB19} are accessible via quasi\-probabilities.
	Even more recently studied phenomena, such as out-of-time-order correlations, are accessible in this manner \cite{YSD18}.

	In general, the notion of what is considered to be nonclassical strongly depends on the chosen classical reference.
	For instance, negativities in the GS distribution do not provide a viable approach for identifying quantumness in anharmonic oscillator systems and specific applications.
	For example, the coherent state can be found to be nonclassical with respect to weak measurement \cite{Ref2a}, which relates to negativities in another (Margenau-Hill) distribution \cite{Ref2b,Ref2c,Ref2d}.
	Thus, a broadly applicable approach to the concept of a quasi\-probability is needed.

	For a general quantum system and a set of classical states, $\hat\rho(c)$, any quantum state can be expanded as \cite{SW18a}
	\begin{equation}
		\label{eq:Coherence}
		\hat \rho=\int \rmd P(c)\, \hat\rho(c)+\hat\rho_{\perp},
	\end{equation}
	which resembles definition \eref{eq:GSrepresentation}.
	In this sense, classical states form a convex hull, described through classical distributions $P\geq0$.
	The linear span of classical state additionally includes states which are nonclassical in this sense, \textit{i.e.} require quasi\-probabilities of the form $P\ngeq0$.
	Beyond this, one can have a residual component $\hat\rho_{\perp}$ in scenarios where the set of classical states $\hat\rho(c)$ does not span the entire space of density operators (note that $\hat\rho_{\perp}=0$ for classical states).

	The relatively simple decomposition \eref{eq:Coherence} and the freedom to choose arbitrary sets of classical states $\hat\rho(c)$, which are typically pure states \cite{P72,P86}, demonstrate the general success of quasi\-probabilities.
	Examples for this can be found in the field of quantum information science (see \cite{NC00} for an introduction) in which nonclassical resources, \textit{i.e.} quantum coherence \cite{SAP17,CG19}, is required to perform certain task which are infeasible in the classical domain \cite{F11,VFGE12}, including metrology applications \cite{AYLLBL19}.

\begin{figure}
	\includegraphics[width=\columnwidth]{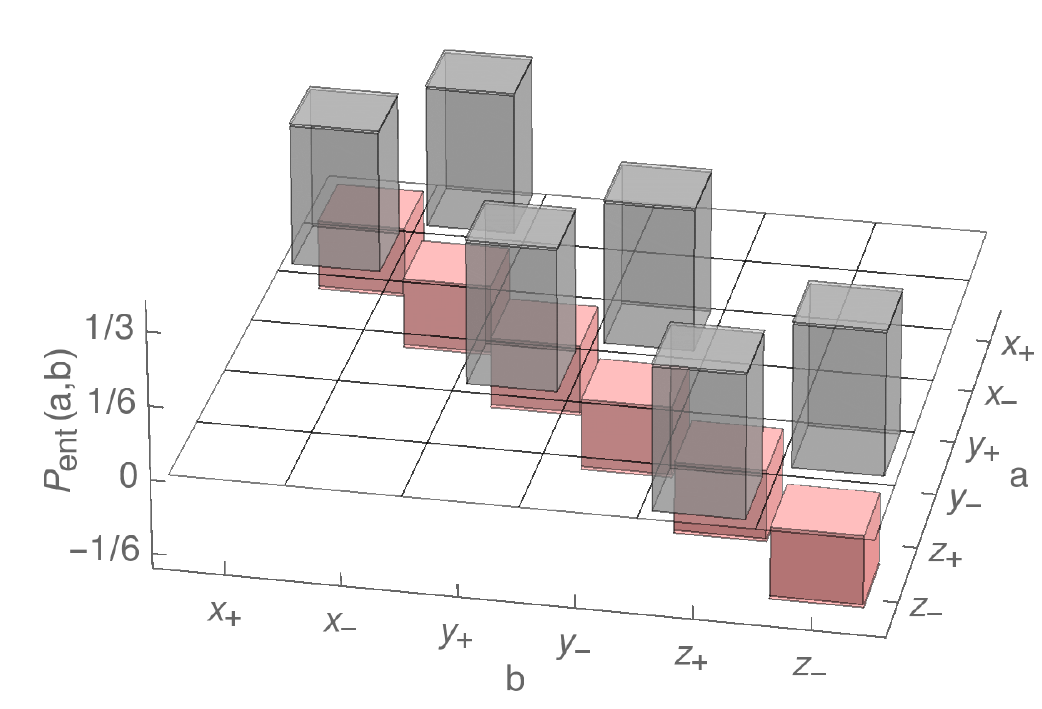}
	\caption{
		Entanglement quasi\-probability distribution $P_\mathrm{ent}$ for a Bell state, $(|0,1\rangle-|1,0\rangle)/\sqrt2$, via tensor products of Pauli-operator eigenstates, $|a\rangle\langle a|\otimes|b\rangle\langle b|$ for $a,b\in\{x_{+},\ldots,z_{-}\}$.
		Inseparability of this state is certified because $P_\mathrm{ent}(a,b)<0$ holds true for elements with $a=b$.
	}\label{fig:Pent}
\end{figure}

	One type of quantumness that is of paramount importance for many applications is entanglement \cite{HHHH09}.
	In this scenario, the classical states $\hat\rho(c)$ are defined as (multimode) tensor-product states, and statistical mixtures thereof define the notion of separable states \cite{W89}.
	Conversely, inseparable states can be represented as pseudo-mixtures \cite{STV98,VT99,GHBELMS02}, also meaning that $\hat\rho_{\perp}=0$ in \eref{eq:Coherence} is always possible.
	Thus, entanglement can be represented via entanglement quasi\-probabilities, which can be constructed for a given density operator by solving nonlinear eigenvalue equations \cite{SV09a,SV09b}.
	See Figure \ref{fig:Pent} for the example of a Bell state.
	The negativities in such a distribution exclusively and unambiguously demonstrate the entanglement of the state under study.
	Other quasi\-probabilities, such as the Wigner function, can do this to a very limited extent only \cite{DMWS06,WFPT17}.
	Entanglement quasi\-probabilities have been applied to a variety of states, as reported in \cite{SV12,BSV17,TBV17}, and generalize to multimode systems \cite{SW18a}. Moreover, entanglement quasi\-probabilities have been recently reconstructed for the first time in an experiment \cite{SMBBS19}, producing a target state like shown in Figure \ref{fig:Pent}.

	Furthermore, and similarly to the bosonic representation, quasi\-probabilities enable us to characterize systems of fermions.
	Again, pioneering work in this direction was done by Glauber and his coworker \cite{CG99}, which was later complemented by representations for Gaussian states \cite{CD06,JRD18}.
	Also quite recently, the application of Grassmann phase-space methods provided a systematic means of studying fermions \cite{E16,DJB16,DJB17}.
	In particular, the study of spin-$1/2$ particles and ensembles thereof vastly benefited from the quasi\-probability framework to study quantum phenomena; for example, see \cite{WBIMH92,P96,TMS01,P12}.

	One has to remark, however, that even states categorized as classical in such qubit systems can still exhibit quantum properties \cite{WB12} because different quantum phenomena violate different classical constraints.
	In fact, it is known that any notion of nonclassicality canonically depends on the choice of the classical reference, which was exemplified for the incompatible concepts of discord and quantum-optical nonclassicality in \cite{FP12}.
	Thus, recent methods attempt to find certain quasi\-probabilistic descriptions with minimal negativities \cite{Z16,BF17}.

	Another operational approach to circumvent the ambiguity of the choice of classical reference states is to formulate quasi\-probabilities for performed measurements instead.
	This description allows one, for example, to efficiently estimate probabilities of measurement outcomes \cite{PWB15}.
	Furthermore, the joint description of detection outcomes from noncommuting observables also requires such negative probabilities to correctly represent the history of sequential measurements \cite{HC16,H17,JRL17,P19}, also demonstrated in a recent experiment \cite{RHLSJLLL18}.
	Moreover, even weak measurements can be characterized via the technique of quasi\-probability distributions \cite{BB10,Ref2f,FLT17}, which further connects to contextuality and the Margenau-Hill distribution \cite{Ref2g}.

	Despite this success of quasi\-probabilities when analyzing the quantumness of different aspects of quantum systems, the general question of how to make a meaningful identification of the classical reference states, \textit{i.e.} the (typically pure) states $\hat\rho(c)$ in \eref{eq:Coherence}, presents a challenging problem.
	For atomic ensembles and angular-momentum-type degrees of freedoms, this question was answered by generalizing coherent states of harmonic oscillators to so-called atomic and angular momentum coherent states \cite{AD71,ACGT72}, both being structurally identical although representing different physical systems.
	The natural phase space of angular-momentum coherent states is the Poincar\'{e} sphere \cite{DAS94,A99} over which the corresponding quasi\-probabilities can be constructed \cite{A81,SW18a}.

	Because angular-momentum states are of such a great importance for different systems, including atomic ensembles, \textit{e.g.} for explaining quantum properties of fluorescence \cite{L84}, much research of quasi\-probabilities in finite-dimensional systems is directed to systems described by employing those states.
	For instance, angular-momentum coherent states allow us to discriminate classical and nonclassical polarization states of light in theory \cite{KM02,L06,GBB08} and experiment, \textit{e.g.} in the recent implementation \cite{SCK17}, for which reconstruction methods have been developed and applied \cite{SBKSL12,MSPGRHKLMS12,KRW14}.
	Moreover, angular-momentum phase-space distributions also enable the identification of entanglement in certain scenarios \cite{A93b,KLLRS02,LLA09,MZHCV15}.

	For other discrete- and continuous-variable systems, physically sensible classical reference states can be also constructed via group-theoretical methods \cite{P72,P86}.
	As a final remark, we may mention that this and similar approaches led to a plethora of methods to devise generalized phase-space methods, \textit{e.g.} in \cite{BM98,LP98,RMG05,RMSG06,KMR06,MSG09,FME10,PB11,RLHL13,TESMN16}, additionally demonstrating the vast interest in quasi\-probability descriptions for identifying different quantum features in a variety of systems.

\section{Visualizing quantum interference}\label{sec:Coherence}

	In the previous section, we provided a broader overview on the successful characterization of quantum phenomena through negative probabilities.
	We demonstrated that such methods are motivated by questions which range from fundamental to practical.
	In the following, we focus on a selection of recently developed quasi\-probability methods and study them in more detail.

	For applications considered in this section, it is useful to recall that each state can be expanded in terms of coherent ones.
	For example, when applying the quantum superposition principle to coherent states, we get pure states which take the form
	\begin{equation}
		\label{eq:CatState}
		|\Psi\rangle=\sum_{k=1}^R\lambda_k|\beta_k\rangle,
	\end{equation}
	using $R$ linearly independent coherent states $|\beta_k\rangle$.
	In fact, the minimal number $R$ of such superpositions of coherent states to represent a given state serves as a nonclassicality measure and can be extended to mixed states as well \cite{GSV12,SV15}.
	Here, we shall focus on the density operator $\hat\varrho=|\Psi\rangle\langle\Psi|$, which expands as
	\begin{equation}\eqalign{
		\label{eq:CatStateVarrho}
		&\hat\varrho
		=\sum_k |\lambda_k|^2|\beta_k\rangle\langle\beta_k|+\sum_{k\neq \tilde k}\lambda_k\lambda_{\tilde k}^\ast |\beta_k\rangle\langle\beta_{\tilde k}|.
	}\end{equation}
	Therein, off-diagonal elements ($k\neq\tilde k$) are a signature of the quantum coherence of this state of light.
	Thus, it is convenient to express those off-diagonal terms in the GS formalism, \textit{cf.} Equation \eref{eq:GSrepresentation}.
	In \ref{sec:Appendix}, it is shown that this can be done when employing complex-valued $\delta$ distributions, meaning that $|\tilde\beta\rangle\langle\beta|=\int \rmd^2\alpha\,P_{|\tilde\beta\rangle\langle\beta|}(\alpha)|\alpha\rangle\langle\alpha|$ and
	\begin{equation}\eqalign{
		\label{eq:POffDiag}
		P_{|\tilde\beta\rangle\langle\beta|}(\alpha)
		&=\langle\beta|\tilde\beta\rangle
		\delta\left(\mathrm{Im}(\alpha)-\frac{\tilde\beta-\beta^\ast}{2\rmi }\right)
		\\&\phantom{=}\times
		\delta\left(\mathrm{Re}(\alpha)-\frac{\tilde\beta+\beta^\ast}{2}\right)
	}\end{equation}
	represent the quantum interference $|\tilde\beta\rangle\langle\beta|$ for $\beta\neq\tilde\beta$, where $\langle\beta|\tilde\beta\rangle=\exp[-|\beta-\tilde\beta|^2/2+\rmi\mathrm{Im}(\beta^\ast\tilde\beta)]$.
	It is also worth mentioning that $P_{|\beta\rangle\langle\tilde\beta|}=[P_{|\tilde\beta\rangle\langle\beta|}]^\ast$ holds true, which, as a consequence, correctly leads to a real-valued GS distribution for Hermitian operators.
	The representation of off-diagonal elements via \eref{eq:POffDiag} becomes quite handy for our following considerations.

\subsection{Regularized GS functions}

	Since the expression \eref{eq:POffDiag} is singular, \textit{i.e.} not a smooth function, an experimental reconstruction of GS functions for states with the corresponding off-diagonal terms cannot be done with finite data.
	Thus, as discussed in Section \ref{sec:Review}, regularization methods have been developed.
	We focus on the approach in \cite{KV10}, which enables one to represent each classical or nonclassical state via a smooth (\textit{cf.} \cite{ASV13}) function that is completely nonnegative or includes negativities, respectively.
	This has been successfully applied to experiments for visualizing nonclassicality via regular phase-space functions, \textit{e.g.} in \cite{KVHS11,KVBZ11} using data obtained from balanced homodyne detection.

	For our purpose, this method can be recast into a convolution as discussed for Equation \eref{eq:Convolution}.
	As an example, we may use the nonnegative and non-Gaussian convolution kernel
	\begin{equation}
		\label{eq:ThomasKernel}
		K(\alpha,\tilde\alpha){=}\frac{w^2}{\pi^2}
		\left[\frac{
			\sin(w\mathrm{Re}[\alpha{-}\tilde\alpha])
		}{
			w\mathrm{Re}[\alpha{-}\tilde\alpha]
		}\right]^2
		\left[\frac{
			\sin(w\mathrm{Im}[\alpha{-}\tilde\alpha])
		}{
			w\mathrm{Im}[\alpha{-}\tilde\alpha]
		}\right]^2\!\!,
	\end{equation}
	which includes an adjustable width parameter $w>0$ \cite{KV10}.
	Then, applying the GS function in \eref{eq:POffDiag} and the integral
	\begin{equation}\eqalign{
		&\phantom{=}\int \rmd u\,\left[\frac{\sin(u)}{u}\right]^2\delta(u-[x+\rmi y])
		\\
		&=\left[\frac{\sin(x)\cosh(y)+\rmi\cos(x)\sinh(y)}{x+\rmi y}\right]^2,
	}\end{equation}
	for real numbers $x$ and $y$, we can express the off-diagonal elements via regularized expressions.

\begin{figure}
	\includegraphics[width=\columnwidth]{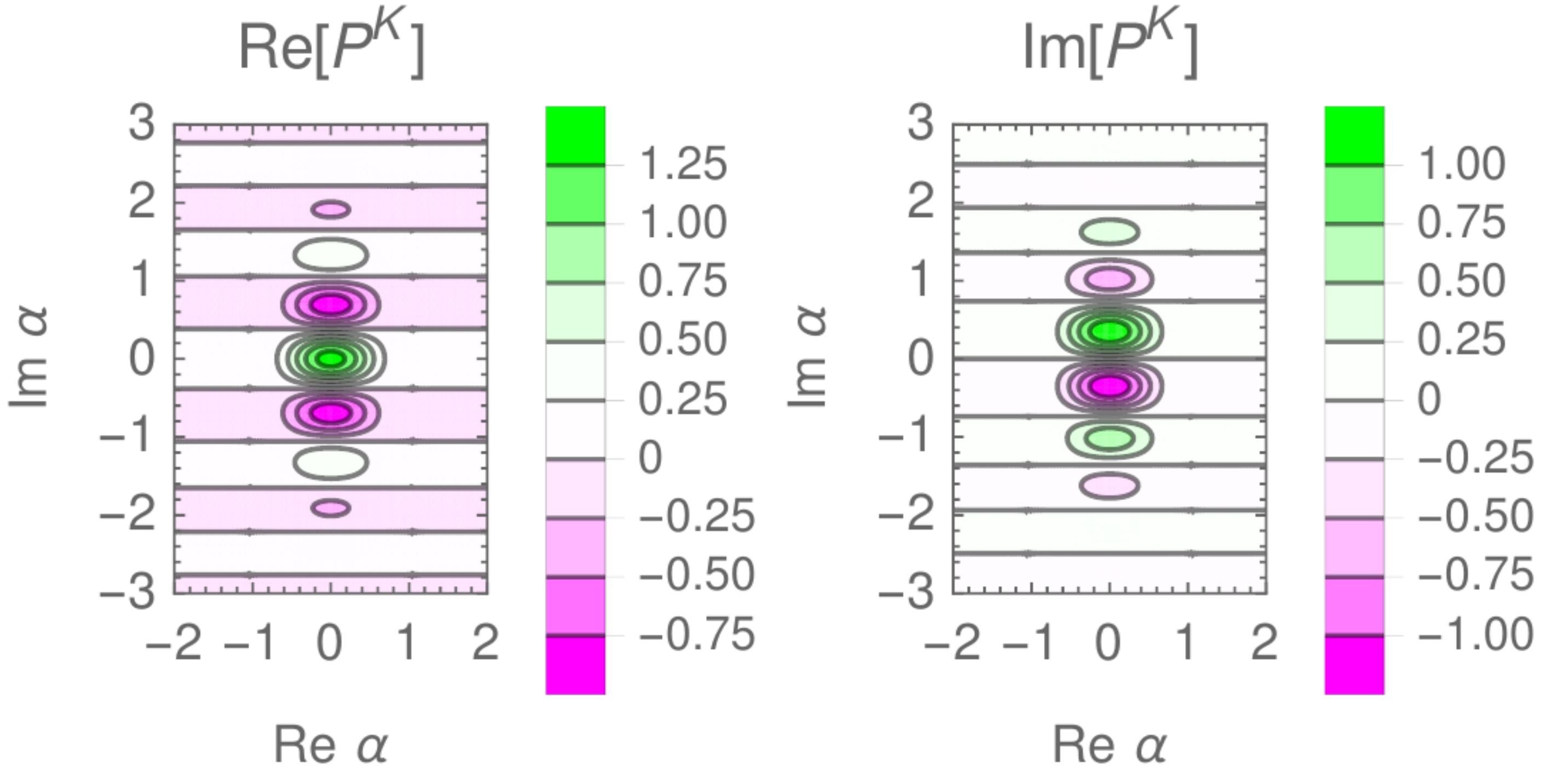}
	\caption{
		Interference GS function \eref{eq:POffDiag}, regularized via the kernel \eref{eq:ThomasKernel} for the width parameter $w=3$.
		The real (left) and imaginary (right) part of $P^K_{|\tilde\beta\rangle\langle\beta|}$ are depicted for coherent amplitudes $\beta=-\tilde\beta=1$.
	}\label{fig:Interference}
\end{figure}

	Based on the analytic formulae above, we can now analyze the impact of quantum superpositions that define the state \eref{eq:CatStateVarrho}.
	In Figure \ref{fig:Interference}, we specifically depict the interference contribution $P^K_{|\tilde\beta\rangle\langle\beta|}$ for $\beta\neq\tilde\beta$.
	Clear quantum interference fringes are visible.
	Those, when added up according to \eref{eq:CatStateVarrho} for formulating the GS function of $\hat\varrho$, certify to the nonclassical behavior from quantum superpositions, as we will see with a later example.
	In addition, while $P_{|\tilde\beta\rangle\langle\beta|}$ is singular, its convoluted version represents a smooth function, as shown in Figure \ref{fig:Interference}.

\subsection{Detector-agnostic GS functions}

	One main issue one encounters when performing measurements for a quantum state reconstruction is that the inner operation of detectors is typically unknown, meaning that only outcomes from the detection process can be recorded.
	Surprisingly, one can overcome this challenge by devising a detector-agnostic approach \cite{SPBTEWLNLGVASW19}, which consists of an unbalanced homodyne configuration and a multiplexed detection \cite{PTKJ96} that uniformly splits a signal into $N=2^d$ output signals, employing $d$ iterations of 50:50 beam splitters.
	In such a scenario, it has been shown that the generating function $G_z$ of the measurement outcome statistics, parametrized throught $z$, is nonnegative for any classical state and even $N$.
	The experimental implementation of this approach can be found in \cite{SPBTEWLNLGVASW19}, where highly sophisticated state-of-the-art detectors are employed.

	Furthermore, one can explicitly express the generating function solely through the state's GS $P$ distribution and the generating function $G_z^\mathrm{(vac)}$, measured for vacuum, as
	\begin{equation}
		\label{eq:DAPS}
		G_z(\alpha)=\int\rmd^2\tilde\alpha\, G^\mathrm{(vac)}_z\left(\alpha-\frac{t}{r}\tilde\alpha\right)P(\tilde\alpha),
	\end{equation}
	where $\alpha$ represents the local oscillator that is mixed on a $|t|^2{:}|r|^2$ beam splitter with the signal, before sending one of the output fields to the multiplexing stage, \textit{cf.} \cite{SPBTEWLNLGVASW19}.
	It is rather interesting to observe that formula \eref{eq:DAPS} resembles the convolution \eref{eq:Convolution}, when identifying $G_z$ and $G_z^\mathrm{(vac)}$ with $P^K$ and $K$, respectively.
	Thus, $G_z$ provides yet another phase-space representation of the state under study.

	When the $N$ detectors used to measure the output signals of the multiplexing stage are on-off detectors, we can directly compute the generating function for vacuum,
	\begin{equation}
		G^\mathrm{(vac)}_z(\alpha)=\left[
			\rme^{-\eta|r|^2|\alpha|^2/N}+z(1-\rme^{-\eta|r|^2|\alpha|^2/N})
		\right]^N,
	\end{equation}
	where $\eta$ is the quantum efficiency; see \cite{LSV15} for details and \cite{BTBSSV18} for the corresponding experimental realization.
	Similarly to our previous regularization, expression \eref{eq:POffDiag} enables us now to compute the result for interference terms of the form $|\tilde\beta\rangle\langle\beta|$.
	Note that for this purpose, we remind ourselves that $|\alpha|^2=(\mathrm{Re}\,\alpha)^2+(\mathrm{Im}\,\alpha)^2$ since we are dealing with complex-valued $\delta$ distributions.
	Taking this into account, we get for the GS function \eref{eq:POffDiag} its detector-agnostic counterpart as
	\begin{equation}
		G_{z,|\tilde\beta\rangle\langle\beta|}(\alpha)
		=\langle\beta|\tilde\beta\rangle
		\left[z+(1-z)\rme^{-\eta r^2 q(\alpha,\beta,\tilde\beta)/N}\right]^N,
	\end{equation}
	where we used the simplifications $t=t^\ast$ and $r=r^\ast$ and the abbreviation
	\begin{equation}\eqalign{
		q(\alpha,\beta,\tilde\beta)
		{=}\left[
			\mathrm{Re}\,\alpha{-}\frac{t}{r}\frac{\tilde\beta{+}\beta^\ast}{2}
		\right]^2
		{+}\left[
			\mathrm{Im}\,\alpha{-}\frac{t}{r}\frac{\tilde\beta{-}\beta^\ast}{2\rmi}
		\right]^2
		\!\!.
	}\end{equation}

\begin{figure}
	\includegraphics[width=\columnwidth]{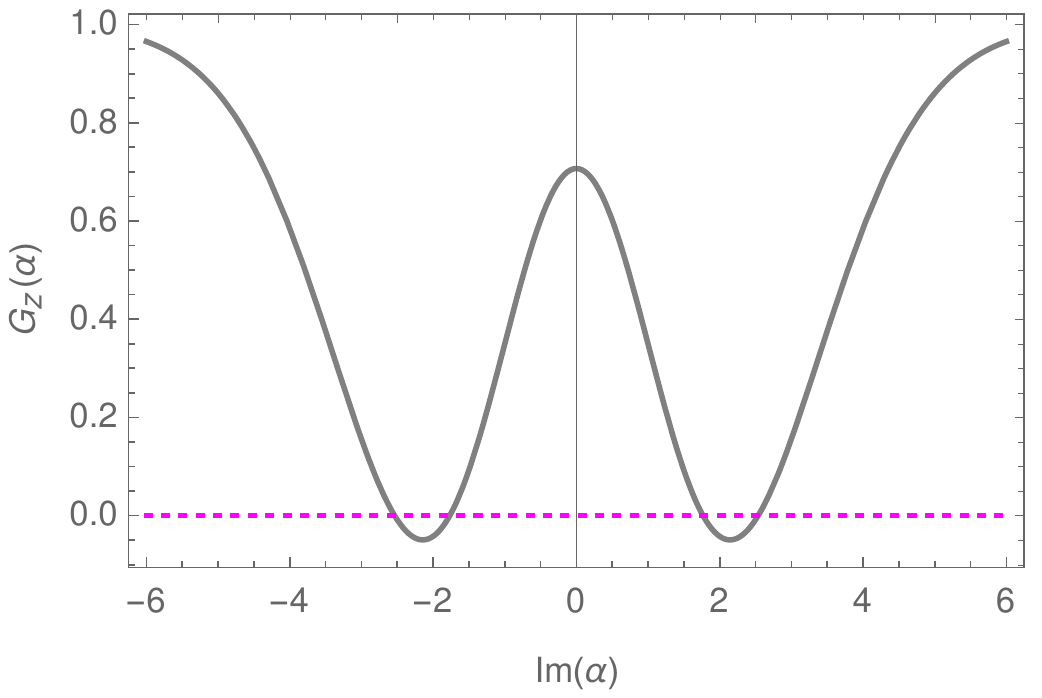}
	\caption{
		Detector-agnostic phase-space representation $G_z$ of an even coherent state, measured with on-off detectors after one multiplexing step, $N=2$.
		A cut along $\mathrm{Re}\,\alpha=0$ is shown.
		Free parameters are chosen as $t=r=1/\sqrt{2}$, $\eta=0.5$, $\beta=1$, and $z=-1$;
		see text for further details.
	}\label{fig:DAPS}
\end{figure}

	Again, we can use the derived analytic expressions to study an example.
	Here, we may consider an even coherent state, $|\Psi\rangle=(|\beta\rangle+|-\beta\rangle)/[2(1+\rme^{-2|\beta|^2})]^{1/2}$.
	Based on our results, we readily get the corresponding phase-space representation for this state as the linear combination $G_z(\alpha)=[G_{z,|\beta\rangle\langle\beta|}(\alpha)+G_{z,|-\beta\rangle\langle\beta|}(\alpha)+G_{z,|\beta\rangle\langle-\beta|}(\alpha)+G_{z,|-\beta\rangle\langle-\beta|}(\alpha)]/[2(1+\rme^{-2|\beta|^2})]$, which is shown in Figure \ref{fig:DAPS}.
	Even for $50\%$ detection losses and a single multiplexing step, $d=1$ (thus $N=2^1$), the depicted phase-space function includes negative contributions which verify the state's nonclassical character.

\subsection{Hybrid GS matrix}

	Our previous quasi\-probability approaches concerned a single optical mode only.
	However, for realizing versatile quantum information platforms, hybrid systems need to be considered \cite{KL12,ANLF15}.
	For instance, light-matter correlations enable us to transfer quantum information from a discrete-variable degree of freedom to a continuous-variable one.
	As an example, we may consider a two-level atom, described through the ground state $|0\rangle$ and the excited state $|1\rangle$, and a superposition state
	\begin{equation}
		\label{eq:HybridState}
		|\Psi\rangle=2^{-1/2}\left[|\beta\rangle\otimes|0\rangle+|-\beta\rangle\otimes|1\rangle\right].
	\end{equation}
	Alternatively, $|0\rangle$ and $|1\rangle$ could also represent the vacuum and single-photon state, respectively, leading to an all-optical state that is entangled in a combination of discrete (photon number) and continuous (coherent amplitudes) variables.
	Then, a projective measurement along $a_0|0\rangle+a_1|1\rangle$ in the second, discrete-variable system yields the superposition state $a_0|\beta\rangle+a_1|-\beta\rangle$, up to a proper normalization.
	This continuous-variable state now carries the quantum information, \textit{i.e.} quantum coherence, of the qubit projector.
	Note that a generalization to qudits can be done straightforwardly.

	In our theoretical and experimental contribution \cite{ASCBZV17}, we have demonstrated that the possibility to transfer quantum information in the manner described above is directly related to a quasi\-probability matrix, initially considered in \cite{WMV97}.
	For the state \eref{eq:HybridState}, for example, we get a quasiproability matrix
	\begin{equation}
		\mathbf{P}(\alpha)=\left[\begin{array}{cc}
			P_{|\beta\rangle\langle\beta|}(\alpha) & P_{|\beta\rangle\langle-\beta|}(\alpha)
			\\
			P_{|-\beta\rangle\langle\beta|}(\alpha) & P_{|-\beta\rangle\langle-\beta|}(\alpha)
		\end{array}\right],
	\end{equation}
	where the rows and columns correspond to the discrete-variable states of the atom, and the entries themselves describe to the corresponding continuous-variable phase-space distributions.
	Then, hybrid nonclassicality is demonstrated when this matrix function is not semidefinite for at least one argument $\alpha$ \cite{ASCBZV17}, \textit{i.e.} $\mathbf{P}\ngeq0$.

	This concept elevates the GS quasi\-probability approach to a matrix-based method.
	Again, convolutions like considered previously are advantageous to obtain smooth distributions;
	see \cite{ASCBZV17} for an application of this concept to experimental data.
	Moreover, the projective measurement to transfer the atomic state of the optical state relates to notions of conditional nonclassicality, which can be compared with joint quantum correlations; see \cite{SAWV17} for an in-depth analysis.

\subsection{Multimode GS function}

	In addition to the composition of one optical mode and a matter system, we may also consider $M$ optical modes.
	Clearly, in this case, one has to consider a multimode GS representation,
	\begin{equation}
		\hat\rho
		=\int \prod_{j=1}^M\rmd^2\alpha^{(j)}\,
		P(\alpha^{(1)},\ldots,\alpha^{(M)})
		\bigotimes_{j=1}^M |\alpha^{(j)}\rangle\langle\alpha^{(j)}|,
	\end{equation}
	for multimode coherent states $|\alpha^{(1)},\ldots,\alpha^{(M)}\rangle=|\alpha^{(1)}\rangle\otimes\cdots\otimes|\alpha^{(M)}\rangle$.
	Singular $P$ distributions can be also regularized through convolutions \cite{ASV13} and the failure to interpret $P$ in terms of a classical probability density reflects the nonclassical nature of the multimode state under study.

	In this context, we may emphasize the relevance of the finding in \eref{eq:POffDiag}, likewise in \ref{sec:Appendix}.
	As we can write the identity $\hat 1^{(j)}=\int \rmd^2\beta^{(j)}|\beta^{(j)}\rangle\langle\beta^{(j)}|/\pi$ for the $j$th mode, we can expand a general multimode density operator as
	\begin{eqnarray}
		\hat\rho&=&\hat 1^{(1)}\otimes\cdots\otimes\hat 1^{(M)}\hat \rho\hat 1^{(1)}\otimes\cdots\otimes\hat 1^{(M)}
		\\\nonumber
		&=&\int \rmd^2\alpha^{(1)}\cdots\rmd^2\alpha^{(M)}\,P(\alpha^{(1)},\ldots,\alpha^{(M)})
		\\\nonumber
		&&\times|\alpha^{(1)},\ldots,\alpha^{(M)}\rangle\langle \alpha^{(1)},\ldots,\alpha^{(M)}|,
	\end{eqnarray}
	where we used \eref{eq:POffDiag} to directly get the GS function as
	\begin{eqnarray}
		\label{eq:GeneralExpansion}
		&&P(\alpha^{(1)},\ldots,\alpha^{(M)})
		\\\nonumber
		&=&\int \rmd^2\beta^{(1)}\cdots\rmd^2\beta^{(M)}\int\rmd^2\tilde\beta^{(1)}\cdots\rmd^2\tilde\beta^{(M)}
		\\\nonumber
		&&\times P_{|\beta^{(1)}\rangle\langle \tilde\beta^{(1)}|}(\alpha^{(1)})\cdots P_{|\beta^{(M)}\rangle\langle \tilde\beta^{(M)}|}(\alpha^{(M)})
		\\\nonumber
		&&\times \frac{\langle\beta^{(1)},\ldots,\beta^{(M)}|\hat\rho|\tilde\beta^{(1)},\ldots,\tilde\beta^{(M)}\rangle}{\pi^{2M}}.
	\end{eqnarray}
	This approach only requires complex $\delta$ distributions and $\langle\beta^{(1)},\ldots,\beta^{(M)}|\hat\rho|\tilde\beta^{(1)},\ldots,\tilde\beta^{(M)}\rangle$, but no other highly singular distributions for representing multimode quantum coherence stemming from off-diagonal elements of the form $|\beta^{(1)},\ldots,\beta^{(M)}\rangle\langle \tilde\beta^{(1)},\ldots,\tilde\beta^{(M)}|$.

	The above decomposition is, for example, interesting for numerical approximations to predict and assess the outcomes of experiments using highly nonclassical states of a multimode radiation field.
	Again, the direct sampling approach in \cite{KVHS11} for balanced homodyning and detector-agnostic methods in \cite{SPBTEWLNLGVASW19} for unbalanced homodyning complement this by allowing to reconstruct regular phase-space functions for multimode systems from experimental data.
	For instance, this enables one to infer the nonclassical interference properties of experimentally produced states.

	Multipartite systems are of particular importance when studying quantum correlations, such as entanglement.
	In fact, one can shown that a nonclassical GS function is necessary (but not sufficient) for entanglement, which can be seen when comparing entanglement quasiproabilities with phase-space quasiproabilities \cite{SV09b}.
	This has interesting consequences for the generation of entangled states.
	For example, it has been studied in great detail that a nonclassical input state is required at the input of a linear optical network to obtain entanglement in the output state; see \cite{X02,KSBK02} for early and \cite{GS16,HBKP16} for more recent contributions in this direction.
	Even more directly, when a state that includes $R$ superpositions of coherent states [\textit{cf.} \eref{eq:CatState}] is mixed in such a network with $M-1$ vacuum states, then the output multipartite Schmidt rank \cite{EB01} is also $R$, allowing for a direct mapping of input optical coherence to output entanglement \cite{VS14}.

	As an example, let us consider a single-mode, odd coherent state which enters such a optical network together with vacuum.
	Suppose the network distributes light in a balanced manner to $M=3$ modes.
	In this scenario, we get an output state of the form
	\begin{equation}
		\label{eq:Tripartite}
		|\Psi_\beta\rangle=\frac{|\beta,\beta,\beta\rangle-|-\beta,-\beta,-\beta\rangle}{\sqrt{2(1-\rme^{-6|\beta|})}}.
	\end{equation}
	Using our result \eref{eq:GeneralExpansion}, we can immediately formulate the six-dimensional (for real-valued phase-space arguments $\mathrm{Re}\,\alpha_j$ and $\mathrm{Im}\,\alpha_j$ and $j\in\{1,2,3\}$) GS distribution of the output state $|\Psi_\beta\rangle\langle \Psi_\beta|$,
	\begin{eqnarray}
		&&P_\beta(\alpha_1,\alpha_2,\alpha_3)
		\\\nonumber
		&=&\frac{
			P_{|\beta\rangle\langle\beta|}(\alpha_1)P_{|\beta\rangle\langle\beta|}(\alpha_2)P_{|\beta\rangle\langle\beta|}(\alpha_3)
		}{2(1-\rme^{-6|\beta|})}
		\\\nonumber
		&&+\frac{
			P_{|-\beta\rangle\langle\beta|}(\alpha_1)P_{|-\beta\rangle\langle\beta|}(\alpha_2)P_{|-\beta\rangle\langle\beta|}(\alpha_3)
		}{2(1-\rme^{-6|\beta|})}
		\\\nonumber
		&&+\frac{
			P_{|\beta\rangle\langle-\beta|}(\alpha_1)P_{|\beta\rangle\langle-\beta|}(\alpha_2)P_{|\beta\rangle\langle-\beta|}(\alpha_3)
		}{2(1-\rme^{-6|\beta|})}
		\\\nonumber
		&&+\frac{
			P_{|-\beta\rangle\langle-\beta|}(\alpha_1)P_{|-\beta\rangle\langle-\beta|}(\alpha_2)P_{|-\beta\rangle\langle-\beta|}(\alpha_3)
		}{2(1-\rme^{-6|\beta|})},
	\end{eqnarray}
	which inherits the negativities from the input GS function of the single-mode odd coherent state.
	Specifically, the second and third summands carry the information about the nonlocal quantum coherence of the state.

	Interestingly, this state can connect in an asymptotic manner two inequivalent classes of tripartite entanglement; namely the GHZ and W state \cite{DVC00}.
	On the one hand, the two employed coherent states become orthogonal for increasing amplitudes, $\langle\beta|-\beta\rangle\to 0$ for $|\beta|\to\infty$.
	Thus, the state \eref{eq:Tripartite} approaches a GHZ state; we may informally write for this case
	\begin{equation}
		|\Psi_\infty\rangle=\frac{|\infty,\infty,\infty\rangle-|-\infty,-\infty,-\infty\rangle}{\sqrt 2}=|\mathrm{GHZ}\rangle,
	\end{equation}
	where $\langle \infty|-\infty\rangle=0$.
	On the other hand, the limiting process $|\beta|\to0$ yields a W state,
	\begin{equation}
		|\Psi_0\rangle=\frac{|0,0,1\rangle+|0,1,0\rangle+|0,0,1\rangle}{\sqrt{3}}=|\mathrm{W}\rangle,
	\end{equation}
	where $|0\rangle$ and $|1\rangle$ denote the vacuum and single-photon state, respectively, and which, for example, can be easily shown with the photon-number basis expansion of $|\Psi_\beta\rangle$.
	Thus, the GS function $P_\beta(\alpha_1,\alpha_2,\alpha_3)$ approximates a W state for small and a GHZ state for large coherent amplitudes $\beta$, representing two distinct families of quantum correlations in multimode systems.


\section{Conclusion}\label{sec:Summay}

	In summary, we provided a comprehensive overview of the methodology of quasi\-probabilities in quantum optics and beyond to identify quantum properties of physical systems.
	We mainly focused on the quantum interference of light as introduced by Glauber and others.
	This included a general assessment of phase-space representations and the discussion of recent advances in this and neighbouring fields of contemporary research.

	Because of the unfavorable behavior of traditional phase-space distributions, we put special emphasis on convolution techniques that enable us to find regular expressions for quasi\-probability distributions which are, in addition, directly experimentally accessible.
	This study included detection schemes and reconstruction methods without full photon-number resolution, phase-space distributions for hybrid continuous- and discrete-variable systems, and instances of multimode light for analyzing quantum correlations.

	It is worth emphasizing that quantum-optical quasi\-probabilities inspired a wide range of novel techniques that concern many applications in the rapidly growing field of quantum information science.
	For example, quasi\-probabilities for entanglement offer a means for studying correlations in multimode systems in terms of negative probabilities, reflecting their quantum correlated nature.
	In this context, we particularly analyzed correlations which stem from the quantum superposition principle applied to coherent states.
	This led to analytical expressions for the phase-space distribution for quantum superpositions, representing a nonclassical state of light.

	Therefore, quasi\-probabilities serve as a versatile means to characterize quantized radiation fields and other quantum systems, including composite ones.
	This paves the way towards a unifying understanding of fundamental quantum effects.
	Further, such quantum phenomena can be exploited for quantum information protocols to perform useful tasks of practical relevance.
	To date, quasi\-probabilities still prove to be a key instrument to visualize quantum phenomena in an intuitive manner for exploring the quantum-classical boundary in theory and experiment.

\ack
	The Integrated Quantum Optics group acknowledges financial support from the Gott\-fried Wilhelm Leibniz-Preis (Grant No. SI1115/3-1).

\appendix
\section{Off-diagonal elements}\label{sec:Appendix}

	For describing off-diagonal elements of the form $|\tilde\beta\rangle\langle\beta|$ in terms of GS distributions, 
	\begin{equation}
		\label{eq:Pdecomp_off-diag}
		|\tilde\beta\rangle\langle\beta|
		=\int \rmd^2\gamma\,P_{|\tilde\beta\rangle\langle\beta|}(\gamma) |\gamma\rangle\langle\gamma|,
	\end{equation}
	we consider the characteristic function which is the Fourier-transformed GS distribution and can be identified with the following, normally ordered expression:
	\begin{equation}\eqalign{
		\Phi_{|\tilde\beta\rangle\langle\beta|}(\tilde\gamma)
		&=\tr(|\tilde\beta\rangle\langle\beta| {:}\exp[\tilde\gamma\hat a^\dag-\tilde\gamma^\ast\hat a]{:})
		\\
		&=\langle\beta|\tilde\beta\rangle e^{\tilde\gamma\beta^\ast-\tilde\gamma^\ast\tilde\beta}.
	}\end{equation}
	The inverse transformation then results in
	\begin{equation}\eqalign{
		\label{eq:InvFT}
		P_{|\tilde\beta\rangle\langle\beta|}(\gamma)
		&=\int \frac{\rmd^2\tilde\gamma}{\pi^2}\, \exp[\tilde\gamma^\ast\gamma-\tilde\gamma\gamma^\ast] \Phi_{|\tilde\beta\rangle\langle\beta|}(\tilde\gamma)
		\\
		&=\langle\beta|\tilde\beta\rangle
		\int\frac{\rmd x}{2\pi}\,\rme^{\rmi x(\gamma-\gamma^\ast+\beta^\ast-\tilde\beta)/(2\rmi)}
		\\&\phantom{=}\times
		\int\frac{\rmd y}{2\pi}\,\rme^{-\rmi y(\gamma+\gamma^\ast-\beta^\ast-\tilde\beta)/2},
	}\end{equation}
	where $\tilde\gamma=(x+\rmi y)/2$ and $x$ and $y$ being real numbers.
	This result can be further expressed via a distribution $g$,
	\begin{equation}\eqalign{
		g(a,b)
		&=\int \frac{\rmd w}{2\pi}\, \rme^{\rmi w(a+ib)}
		=\int \frac{\rmd w}{2\pi}\, \rme^{-\rmi w(a+ib)}
		\\
		&=\sum_{n=0}^\infty \frac{(\rmi b\partial_a)^n}{n!}\int \frac{\rmd w}{2\pi}\, \rme^{\rmi wa}
		=\rme^{\rmi b\partial_a}\delta(a),
	}\end{equation}
	with real-valued parameters $a$ and $b$.
	In order to understand how this distribution acts on functions, we compute
	\begin{equation}
		\int \rmd a\, g(a,b) f(a)= \sum_{n=0}^\infty \frac{\partial_a^n f(a)|_{a=0}}{n!} (-\rmi b)^n=f(-\rmi b),
	\end{equation}
	using the Taylor expansion of the function $f$.
	Consequently, the distribution $g$ is the complex extension of the Dirac distribution, $g(a,b)=\delta(a+\rmi b)$, and we find
	\begin{equation}\eqalign{
	\label{eq:off-diag_dist}
		P_{|\tilde\beta\rangle\langle\beta|}(\gamma)
		&=\langle\beta|\tilde\beta\rangle
		\delta\left(\mathrm{Im}(\gamma)-\frac{\tilde\beta-\beta^\ast}{2\rmi }\right)
		\\&\phantom{=}\times
		\delta\left(\mathrm{Re}(\gamma)-\frac{\tilde\beta+\beta^\ast}{2}\right).
	}\end{equation}


\end{document}